\documentclass[preprint,review,12pt]{elsarticle}

\usepackage{lineno,hyperref}
\modulolinenumbers[5]

\journal{Acta Materialia}

\usepackage{graphicx}% Include figure files
\usepackage{color}
\usepackage{booktabs}
\usepackage{tabularx, booktabs, makecell, caption}

\usepackage{amssymb}
\usepackage{amsmath}

\usepackage{gensymb}
\usepackage{textcomp}
\usepackage[version=4]{mhchem}

\begin{document}
\begin{frontmatter}

\title{Evidence of nickel ions dimerization in \ce{NiWO4} and \ce{NiWO_4-ZnWO_4} solid solutions probed by EXAFS spectroscopy and reverse Monte Carlo simulations}

\author[ISSP]{Georgijs Bakradze\corref{mycorrespondingauthor}}
\cortext[mycorrespondingauthor]{Corresponding author}
\ead{georgijs.bakradze@cfi.lu.lv}

\author[ISSP]{Aleksandr Kalinko}

\author[ISSP]{Alexei Kuzmin}

\address[ISSP]{Institute of Solid State Physics, University of Latvia, 8 Kengaraga street, Riga LV-1063, Latvia}

\begin{abstract}
The existence of exchange-coupled \ce{Ni^{2+}} ions -- the so-called magnetic dimers -- in wolframite-type \ce{NiWO4} and \ce{Zn_cNi_{1-c}WO_4} solid solutions with high nickel content was discovered by X-ray absorption spectroscopy combined with reverse Monte Carlo (RMC) simulations. Temperature- (10--300~K) and composition-dependent x-ray absorption spectra were measured at the Ni K-edge, Zn K-edge, and W L$_3$-edge of microcrystalline \ce{NiWO4}, \ce{Zn_cNi_{1-c}WO_4} and \ce{ZnWO4}. Structural models were obtained from simultaneous analysis of the extended x-ray absorption fine structure (EXAFS) spectra at three metal absorption edges using RMC simulations. The obtained radial distribution functions for different atomic pairs made it possible to trace in detail the changes in the local environment of metal ions and the effect of thermal disorder. Dimerization of \ce{Ni^{2+}} ions within quasi-one-dimensional zigzag chains of \ce{[NiO6]} octahedra was evidenced in \ce{NiWO4} in the whole studied temperature range. It manifests itself as the splitting of the Ni--Ni radial distribution function into two separate peaks. The effect is further preserved in solid solutions \ce{Zn_cNi_{1-c}WO_4} for $c \leq 0.6$, which is related to the probability to find two Ni$^{2+}$ ions in neighbouring positions. 
\end{abstract}


\end{frontmatter}

%\linenumbers

\newpage

\section{Introduction}\label{s:intro}
X-ray absorption spectroscopy (XAS) is an experimental technique which -- under favorable conditions -- exhibits an exceptional sensitivity to minute changes of local structure. Magnetostriction-induced variation of interatomic distances in iron-cobalt thin films \cite{Pettifer2005}, electric field-induced structural changes in oxygen-deficient ceria \cite{Korobko2015}, and isotopic effect on atomic vibrations and nearest-neighbour distances in crystalline germanium \cite{Purans2008ge} were successfully studied in the past by XAS. 
Recent developments in advanced data analysis based on the reverse Monte Carlo (RMC) method solved many problems inherent to conventional approaches, thus, enabling one to use XAS for the extraction of reliable structural information such as radial distribution functions (RDFs) or mean-squared relative displacement (MSRD) in complex materials \cite{Nemeth2012, Timoshenko2012rmc, Timoshenko2014zno, Timoshenko2017a, Timoshenko2017c, Levin2017, DiCicco2018, Kraynis2019, Pethes2019}. This unique opportunity can be employed to probe fine structural effects beyond the first coordination shell in the region of the medium-range order, where coordination shells overlap and the contributions from many-atom distribution functions -- known as multiple-scattering (MS) effects -- become important in XAS \cite{Rehr2000, Natoli2003, Rehr2009}. For example, tungstates with a wolframite-type structure represent an interesting class of materials \cite{timoshenko2014cuwo4,timoshenko2015mncowo4,timoshenko2016cowo4} where new opportunities in XAS data analysis can be readily exploited. 

\begin{figure}[t]
	\centering
	\includegraphics[width=0.8\textwidth]{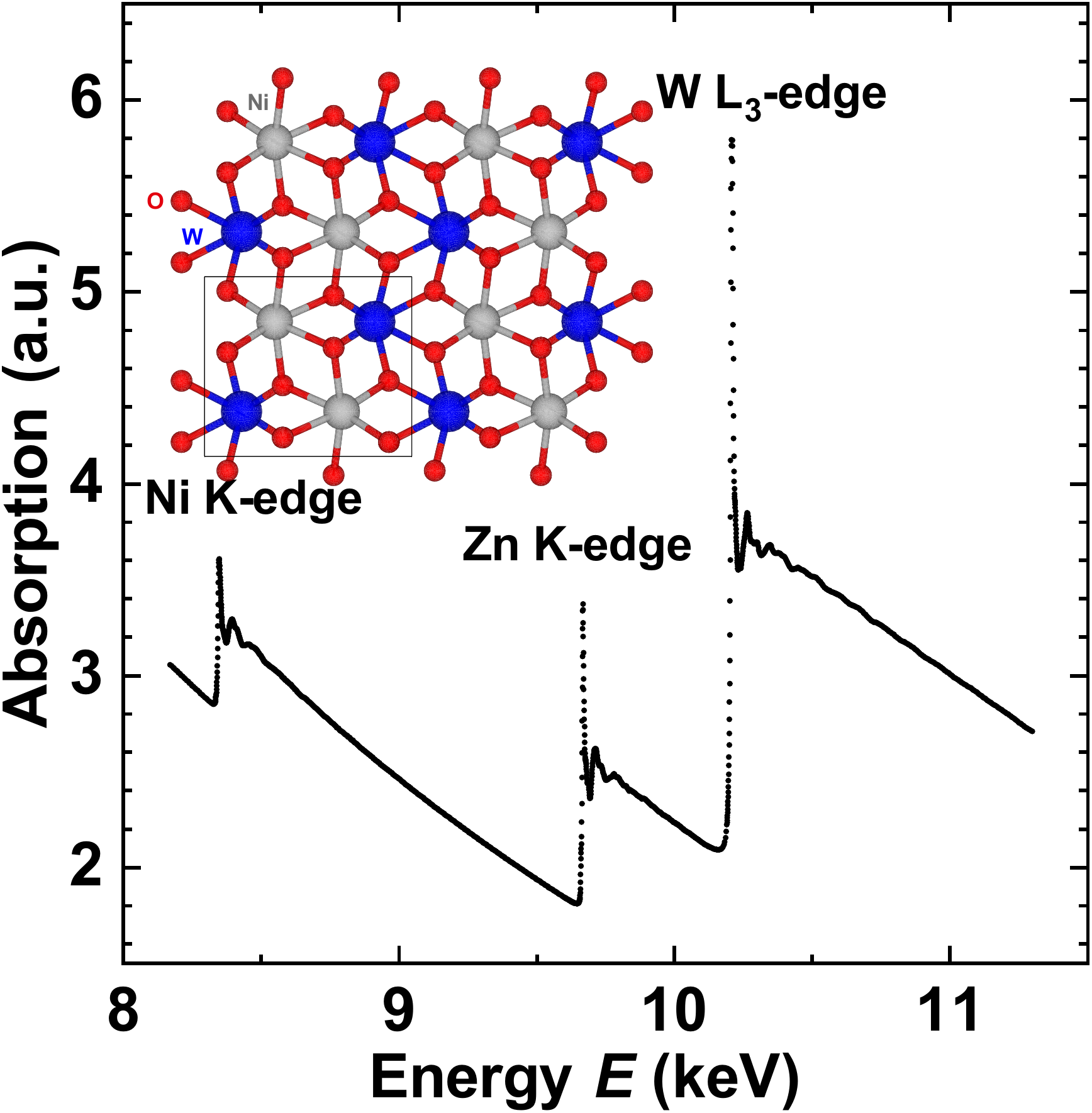}
	\caption{X-ray absorption spectrum of microcrystalline \ce{Zn_{0.6}Ni_{0.4}WO4} spanning across the range of the Ni and Zn K-edges and W L$_3$-edge at 300~K. Inset shows a single layer of \ce{NiWO4} crystal structure oriented so that the (100)-plane is within the plane of the drawing, the directions [010] and [001] are parallel to the abscissa and ordinate, respectively.}
	\label{fig1}
\end{figure}

Pure \ce{NiWO4} and \ce{ZnWO4} crystallize in the structural type of wolframite with the monoclinic crystal system (space group $P2/c$ (13), $Z$=2) \cite{Keeling1957, Filipenko1968, Trots2009}. The wolframite structure can be thought of as a hexagonal closest-packed array of oxygen anions with the tungsten and metal cations in the octahedral voids, forming infinite zigzag chains along the [001]-direction of the crystal; in the perpendicular [100]-direction short chains of \ce{[MO6]} octahedra form a layer which alternates with a layer of edge-joined \ce{[WO6]} octahedra (see the inset in Fig.\ \ref{fig1}). Each chain of \ce{[WO6]} octahedra is attached by common corners to four chains of \ce{[MO6]} octahedra and vice versa. The \ce{[MO6]} octahedra are slightly distorted, whereas \ce{[WO6]} octahedra are strongly distorted with tungsten ions being located off-centre due to the second-order Jahn-Teller effect induced by the \ce{W^{6+}}(5d$^0$) electronic configuration \cite{KUNZ1995}. The lattice dynamics of \ce{NiWO4} and \ce{ZnWO4} have been extensively studied in the past by Raman and infrared spectroscopies in \cite{Oliveira2008, Liu1988, Wang1992, Fomichev1994, Kuzmin2001a, Errandonea2008, Kuzmin2011niwo4}.

Magnetic properties of \ce{NiWO4} have drawn considerable attention in the past: it is known that below the N\'eel temperature ($T_\text{N}$=60--67~K \cite{Wilkinson1977,Prosnikov2017, Liu2017}), the antiferromagnetic AF1-type ordering of spins \cite{Weitzel1970,Weitzel1976} prevails in \ce{NiWO4}. The quasi-one-dimensional zigzag chains of \ce{[NiO6]} octahedra with magnetic Ni$^{2+}$(3d$^8$) ions may be considered as a one-dimensional $S=+1$ system. At the same time, \ce{ZnWO4} is a diamagnetic material, so one can expect that the dilution of nickel ions with zinc ions will affect the magnetic ordering and local interactions in \ce{Zn_cNi_{1-c}WO_4} solid solutions, thus, influencing the local structure of the compound. Close values of the ionic sizes (0.69~\AA\ for Ni$^{2+}$ and 0.74~\AA\ for Zn$^{2+}$, respectively \cite{Shannon1976}) and electronegativity values (1.91 for Ni and 1.65 for Zn) favour the formation of a continuous series of solid solutions between isomorphous \ce{NiWO4} and \ce{ZnWO4}. The properties and possible applications of these solid solutions were seldom studied in the past: the long-range structure, optical, electrical, and magnetic properties of \ce{Zn_cNi_{1-c}WO_4} solid solutions were characterized in \cite{Oliveira2008, Karmakar2020, Kalinko2011a, Huang2015}, and their possible use as a yellow pigment has been proposed in \cite{Oliveira2008}.

Besides \ce{NiWO4}, several other tungstates (\ce{MnWO4}, \ce{CoWO4} and \ce{CuWO4} \cite{Weitzel1970,MnWO4}) exhibit long-range antiferromagnetic ordering at low temperatures. Our previous XAS studies of \ce{CuWO4} \cite{timoshenko2014cuwo4} and \ce{CoWO4} \cite{timoshenko2016cowo4} based on the detailed analysis of metal--oxygen and metal--metal RDFs obtained by RMC simulations agree well with the known diffraction data, suggesting the existence of paired 3d ions with alternating interatomic distances in \ce{CuWO4} and their absence in \ce{CoWO4}. 

In this study, we have applied XAS to the case of \ce{NiWO4} with the crystallographic structure similar to that of \ce{CoWO4}. According to previous diffraction studies \cite{Keeling1957,Wilkinson1977,Weitzel1970}, in \ce{NiWO4} the alternating Ni--Ni distances within the zigzag chains formed by \ce{[NiO6]} octahedra are absent. Nevertheless, we have found that an accurate analysis of the extended X-ray absorption fine structure (EXAFS) based on RMC simulations gives strong evidence that local structural deviations from the average crystallographic structure probed by diffraction exist in pure \ce{NiWO4} in a wide temperature range (10--300 K) as well as upon dilution of \ce{NiWO4} by diamagnetic zinc ions down to $c=0.6$ in \ce{Zn_cNi_{1-c}WO_4} solid solutions at 300~K. The observed peculiarities of the local structure appear as a pairing of nickel ions along the [001] direction, being qualitatively similar to the case of Cu--Cu dimers in \ce{CuWO4} \cite{Lake1996,Lake1997,Koo2001}.

\section{Experimental and data analysis}\label{s:exper}
Pure tungstates (\ce{NiWO4} and \ce{ZnWO4}) and \ce{Zn_cNi_{1-c}WO4} solid solutions were produced by the co-precipitation method. Precipitates were obtained by mixing proper amounts of aqueous solutions of \ce{Na2WO4.2H2O} and \ce{ZnSO4.7H2O} and/or \ce{Ni(NO3)2.6H2O} at room temperature (20~$\celsius$) and pH=8 followed by annealing at 800~$\celsius$ in the air (see \cite{Kalinko2011a} and \cite{Bakradze2020} for details).

XAS measurements were performed in transmission mode at the HASYLAB DESY C1 bending-magnet beamline \cite{DORISC} in the temperature range from 10~K to 300~K at the Ni K-edge (8333~eV), Zn K-edge (9662~eV) and W L$_3$-edge (10207~eV). The storage ring DORIS III operated at $E$=4.44~GeV and $I_\text{max}$=140~mA. The x-ray beam intensity was measured by two ionization chambers filled with argon and krypton gases. The higher-order harmonics were effectively eliminated by detuning the double-crystal monochromator Si(111) to 60\% of the rocking curve maximum, using the beam-stabilization feedback control. The Oxford Instruments LHe flow through cryostat was used to maintain the preset sample temperature. The powder samples were deposited on the Millipore filters and fixed by Scotch tape. The sample weight was chosen to result in the absorption edge jump close to 1.0. Three absorption edges were recorded in a single wide-range scan. Note that the EXAFS signal at the Zn K-edge can only be collected up to about 10~\AA$^{-1}$ due to the vicinity of the W L$_3$-edge (Fig.\ \ref{fig1}). 

The EXAFS spectra $\chi(k)k^2$ were extracted from the total absorption spectra using conventional procedure \cite{Kuzmin2014} as a function of the photoelectron wavenumber $k = \sqrt{(2 m_\text{e}/\hbar^2) (E - E_0)}$, where $E$ is the photon energy, $E_0$ is the minimal energy required to excite a core electron from the respective shell, $m_\text{e}$ is the electron mass, and $\hbar$ is the reduced Planck’s constant. Comparison of several EXAFS measurements -- performed for each sample -- indicates that the statistical noise in the EXAFS signal was much lower than the systematic uncertainty of data processing. To analyze the EXAFS spectra, we employed the RMC and evolutionary algorithm (EA) technique simulation scheme as implemented in the EvAX software package, described elsewhere \cite{Timoshenko2012rmc,Timoshenko2014rmc}. Our structural model was a supercell of 4$a$$\times$4$b$$\times$4$c$ size ($a$, $b$, $c$ are the lattice parameters as-determined from XRD studies \cite{Keeling1957, Filipenko1968}). The coordinates of atoms are changed in a random iterative process, aimed to minimize the difference between the Morlet wavelet transforms (WTs) of the experimental and configuration-averaged (CA) EXAFS spectra. The atomic configurations of solid solutions necessary for the RMC simulation procedure were generated by a random substitution of a part of nickel atoms with zinc in the supercell in the required proportion. A separate fit parameter study showed that clustering of Ni cations in the Zn-rich phase (or Zn cations in the Ni-rich phase) did not result in better RMC simulations. At the same time, at low Ni concentration, one cannot prove the presence or absence of Ni-Ni dimers, because their relative contribution into the total EXAFS spectrum is very small and falls beyond the sensitivity of our method. 

Theoretical CA-EXAFS spectra were calculated at each iteration using the ab-initio real-space MS FEFF8.5L code \cite{Ankudinov1998} for the given structure model. The scattering amplitude and phase shift functions required for the simulation were calculated by the ab initio multiple-scattering FEFF8.5L code using the complex exchange-correlation Hedin-Lundqvist potential \cite{Rehr2000,Ankudinov1998}. The calculations were performed based on the crystallographic structure of \ce{ZnWO4} \cite{Filipenko1968}, \ce{NiWO4} \cite{Keeling1957} or random substitutional solid solution with the respective concentration considering an 8~\AA\ cluster around the absorbing atom (Ni, Zn and W). Calculations of the cluster potentials were done in the muffin-tin (MT) self-consistent-field approximation using the default values of the MT radii as implemented in the FEFF8.5L code \cite{Ankudinov1998}. 

The best-fits of EXAFS spectra were performed in the WT ($k$,$R$)-space \cite{Timoshenko2009wavelet}. In $k$-space, the ranges were selected equal to 3.0--16.0~\AA$^{-1}$ (for Ni and W) and 3.0--10.0~\AA$^{-1}$ (for Zn). In $R$-space, the ranges were set to 1.0--5.5~\AA\ for the Ni and Zn K-edges and 1.0--6.5~\AA\ for the W L$_3$-edge. The values of $E_0$ were set to 8333.0~eV, 9662.0~eV, and 10210.5~eV for the Ni K-edge, Zn K-edge, and W L$_3$-edge, respectively, to have the best match between the experimental and calculated EXAFS spectra.

\begin{figure*}[t]
	\centering
	\includegraphics[width=0.9\textwidth]{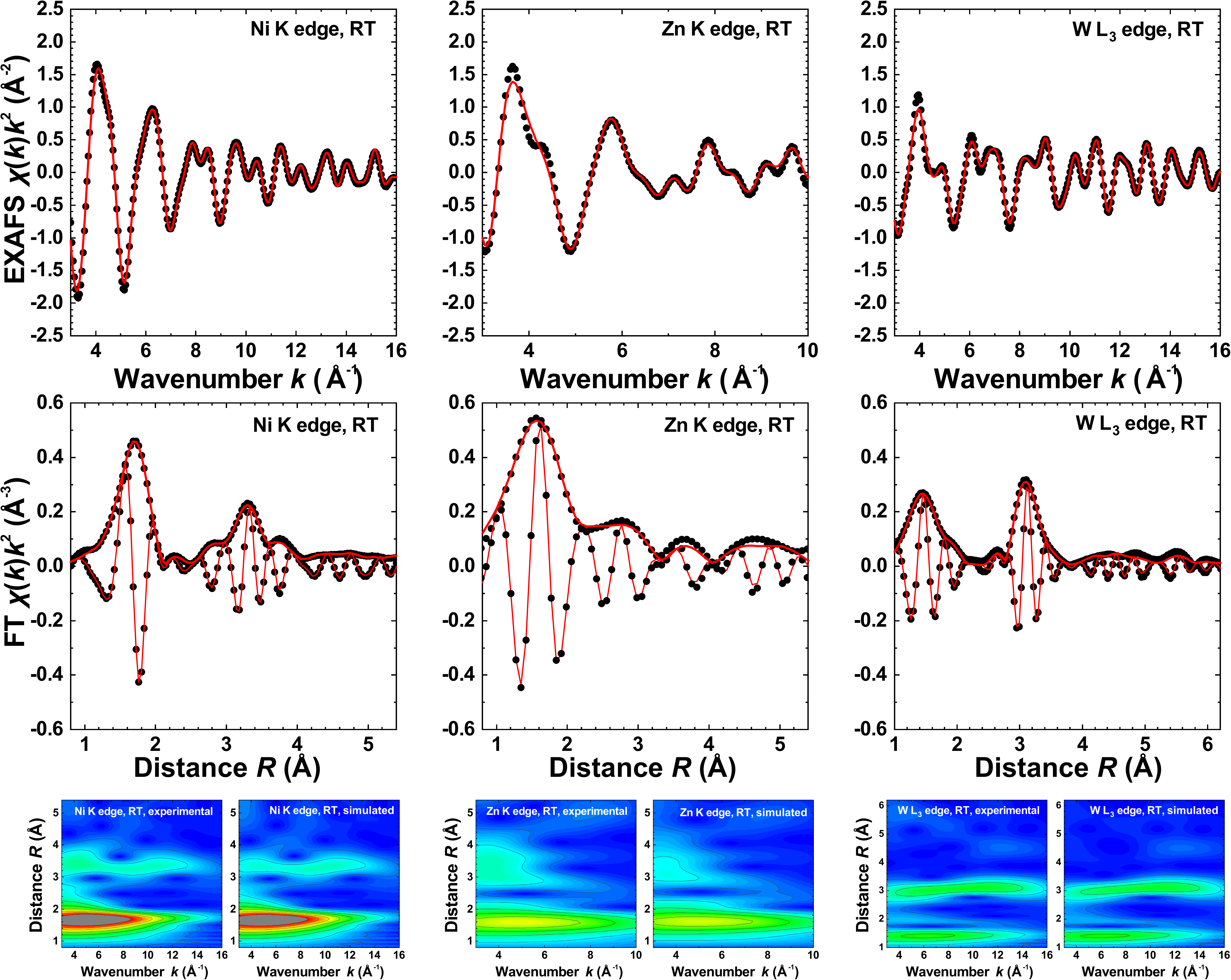}	
	\caption{Experimental (black dots) and simulated (solid lines) extended X-ray absorption fine structure (EXAFS) spectra $\chi(k)k^2$ of \ce{Zn_{0.6}Ni_{0.4}WO4} solid solution at the Ni and Zn K-edges and W L$_3$-edge (top row), their Fourier transforms (FTs) (middle row) and wavelet transforms (bottom row) at 300~K. Both modulus and imaginary parts are shown in FTs. Examples of wavelet transform of the EXAFS spectra at the Ni and Zn K-edges and W L$_3$-edge in $k$-$R$ space at RT are shown in the lower row. Experimental data (left panel) and simulated data (right panel) are shown. See text for details.}
	\label{fig2}
\end{figure*}

\begin{figure*}[t]
	\centering
	\includegraphics[width=0.9\textwidth]{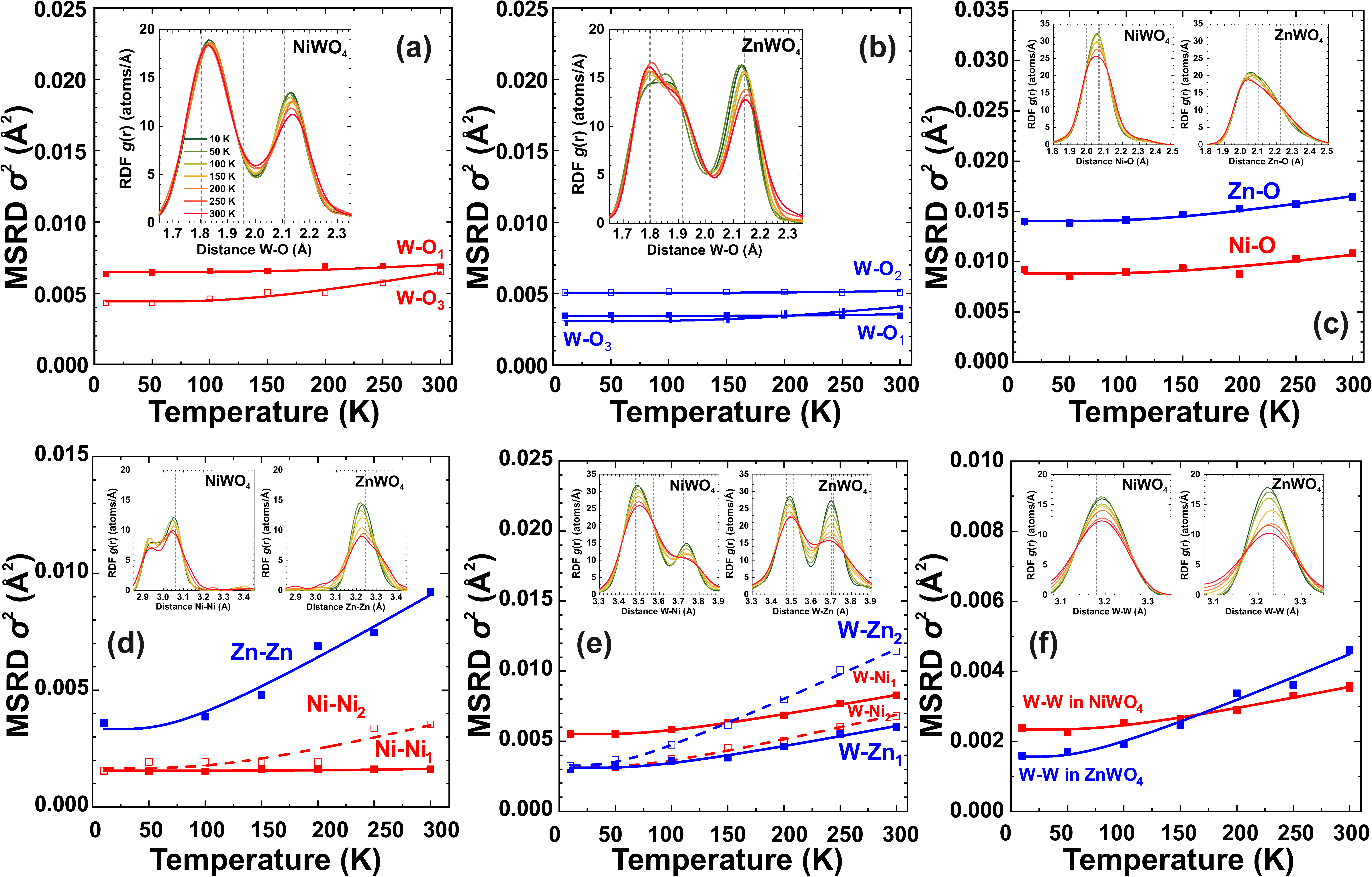}
	\caption{The mean-square relative displacements (MSRDs) $\sigma^2$ for W--O (a,b), Zn--O and Ni--O (c), Zn--Zn and Ni--Ni (d), W--Zn and W--Ni (e), and W--W (f) atomic pairs in microcrystalline \ce{NiWO4} and \ce{ZnWO4} as a function of temperature in the range of 10--300 K. The indices indicate different groups of atoms located in the same coordination shell.  Lines in the main plots are the $\sigma^2(T)$ curves fitted to the correlated Einstein model \protect\cite{Sevillano1979}. Temperature-dependent radial distribution functions (RDFs) $g(r)$ for metal--oxygen and metal--metal atomic pairs are shown in the insets. The gray vertical lines indicate the respective bond lengths as determined from XRD at 300~K. See text for details.}
	\label{fig3}
\end{figure*}

For each sample and for each simulation run, a single structural model was determined using RMC simulations from EXAFS spectra measured at three metal absorption edges. The atomic configurations obtained from RMC simulations were used to calculate the partial RDFs for metal--oxygen and metal--metal atomic pairs (Fig.\ \ref{fig3}). The RDFs were averaged over at least five simulations performed using different random seed numbers or using different random atomic configurations of solid solutions and smoothed using Fourier basis functions with the algorithm as implemented in Origin software (2015, Beta3).

The obtained RDF curves were decomposed into a set of Gaussian functions to determine the total MSRDs, $\sigma^2$, for metal--oxygen and metal--metal atomic pairs which consists of the static and thermal disorder contributions. The characteristic Einstein frequencies, $\omega$, and bond-stretching constants, $\kappa$, were obtained (Table\ \ref{tab1}) by fitting the temperature-dependent part of the MSRDs to the correlated Einstein model \cite{Sevillano1979} using the Einstein frequency and offset as the only two fitting parameters. Since in disordered substitutional solid solutions, alloying atoms occupy sites of a regular lattice -- thus, the atom type, atomic mass and force constants randomly vary from site to site -- the obtained values of the characteristic Einstein frequencies and bond-stretching force constants should be considered as effective local values. The relative fitting error for the characteristic Einstein frequencies for the most atomic pairs reported in Table\ \ref{tab1} was always lower than 5 \% (some exceptions being the pairs W--O$_2$ in \ce{ZnWO4}, Ni--Ni$_1$ and Ni--Ni$_2$; for these atomic pairs the fitting uncertainty was equal to 11.84 \%, 13.7 \%, and 7.28 \%, respectively).

\begin{table}[t]
	\centering
	\caption{The characteristic Einstein frequencies ($\omega$) and bond-stretching constants ($\kappa$) for the atomic pairs W--O, M--O, W--M, M--M and W--W (M = Zn, Ni) in \ce{NiWO4} and \ce{ZnWO4}. O$_1$, O$_2$, O$_3$ and M$_1$, M$_2$ are three and two nearest groups of oxygen and metal atoms located in the first and second coordination shells, respectively. } 
	\begin{tabular}[t]{lcccc}
		\hline
		& \multicolumn{2}{c}{\makecell{\ce{NiWO4}}} & \multicolumn{2}{c}{\makecell{\ce{ZnWO4}}}\\
		\hline
		& $\omega$ (THz) & $\kappa$ (N/m) & $\omega$ (THz) & $\kappa$ (N/m) \\
		\hline
		W--O$_1$ & 91.3  & 122.7 & 131.0 & 252.3 \\
		W--O$_2$ & -     & -     & 131.0 & 252.0 \\
		W--O$_3$ & 59.6  & 52.3  & 75.0  & 82.7  \\
		M--O$_1$ & 64.6  & 52.5  & 58.4  & 43.8  \\
		W--M$_1$ & 35.2  & 55.1  & 33.2  & 53.1  \\
		W--M$_2$ & 31.4  & 43.9  & 21.7  & 22.6  \\
		M--M$_1$ & 123.3 & 445.9 & 29.8  & 29.0  \\
		M--M$_2$ & 48.5  & 68.9  & --    & --    \\
		W--W     & 36.7  & 123.7 & 25.6  & 60.4  \\
		\hline
	\end{tabular}\label{tab1}
\end{table}

\section{Results and discussion}\label{s:results}
Figure\ \ref{fig2} shows the experimental and RMC/EA calculated EXAFS spectra and their Fourier transforms (FTs) and WTs for \ce{Zn_{0.6}Ni_{0.4}WO4} solid solution at 300~K (data for pure components at 10~K and 300~K can be seen in Figs.~\ref{figa1} and \ref{figa2}, respectively). It is important that a good agreement between the experimental and calculated EXAFS data simultaneously at two (pure components) or three absorption edges (solid solutions) was obtained for each sample in $k$- and $R$-space within a single structural model. The wide $k$ and $R$ range of experimental EXAFS spectra used in the analysis allowed us to probe equally well the contributions from light (oxygen) and heavy (tungsten) atoms, which let us determine reliable structural models for different sample compositions and temperatures. Note that the peak positions in FTs and WTs differ from the crystallographic values due to the phase shift present in EXAFS.

\begin{figure*}[t]
	\centering
	\includegraphics[width=0.85\textwidth]{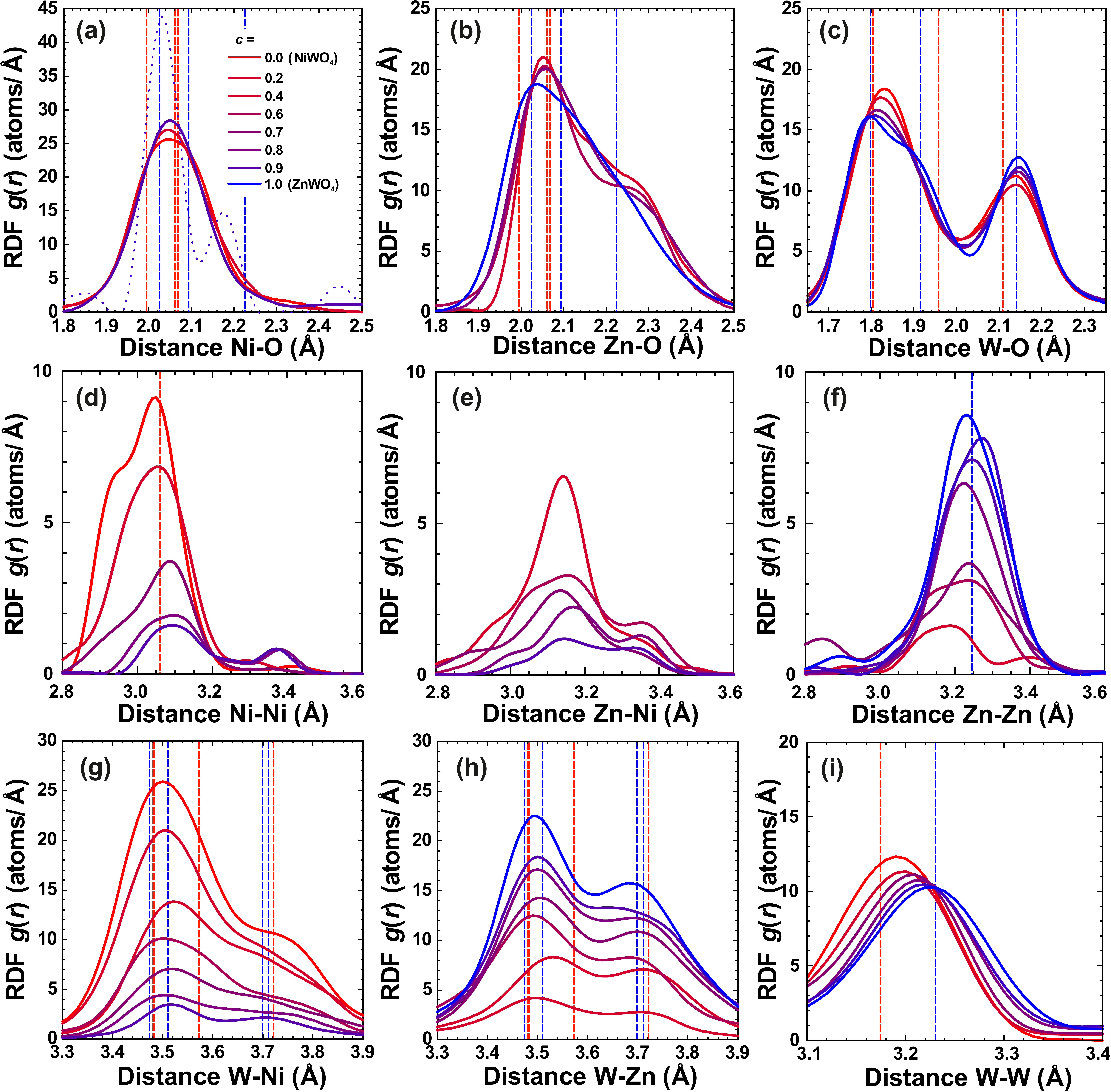}
	\caption{Radial distribution functions (RDFs) $g_\text{M-O}(r)$ (a, b), $g_\text{W-O}(r)$ (c), $g_\text{M-M}(r)$ (d-f), $g_\text{W-M}(r)$ (g, h) (M = Ni, Zn) and $g_\text{W-W}(r)$ (i) in microcrystalline \ce{Zn_cNi_{1-c}WO4} solid solutions at 300~K. The red and blue vertical lines indicate the respective bond lengths in \ce{NiWO4} and \ce{ZnWO4}, respectively, as-determined by XRD at 300~K. See text for details.}
	\label{fig4}
\end{figure*}

The coordinates of atoms in the final atomic configuration obtained in the RMC simulation were used to calculate a set of the partial RDFs for different atom pairs. Temperature-dependent RDFs $g_\text{W-O}(r)$, $g_\text{Ni-Ni}(r)$, $g_\text{Zn-Zn}(r)$, $g_\text{W-Ni}(r)$, and $g_\text{W-Zn}r)$ in pure tungstates are shown in the insets in Fig.\ \ref{fig3} in the temperature range 10--300~K. The RDFs $g_\text{W-O}(r)$ (Fig.\ \ref{fig3}(a,b)) for the first coordination shell of tungsten are qualitatively close to that obtained by us in \cite{Bakradze2020} using the regularization method. However, structural models from the RMC simulations, which take into account contributions from distant coordination shells and several absorption edges, predict slightly narrower $g_\text{W-O}(r)$ distributions than in \cite{Bakradze2020} due to the method used.  It is also noticeable that partial RDFs in Fig.\ \ref{fig3} demonstrate different temperature dependencies. The strongest changes of the RDF shape occur for Zn--Zn, W--Zn$_2$ and W--W atom pairs in \ce{ZnWO4}, whereas most metal--oxygen and Ni--Ni$_1$ atom pairs show the weakest temperature dependence. The composition dependence of RDFs in \ce{Zn_{c}Ni_{1-c}WO4} solid solutions at 300~K is shown in Fig.\ \ref{fig4}. Here the main effect is observed in the outer coordination shells due to the substitution of nickel ions by zinc.

\subsection{Temperature-induced effects in pure \ce{NiWO4} and \ce{ZnWO4}}\label{ss:temp} 
A temperature increase leads to a larger amplitude of atomic vibrations and an expansion of the tungstate lattice. In the studied temperature range, the thermal expansion of the lattice parameters is about 0.01~\AA\ for \ce{ZnWO4}, as determined by synchrotron and neutron powder diffraction in the range of 3--300~K \cite{Trots2009}. Thus, the thermal expansion is weak and, in any case, is smaller than the accuracy (about $\pm$0.02~\AA) of the interatomic distance determination from our EXAFS data. At the same time, an increase in thermal disorder results in a broadening of the RDF peaks, and, thus, the MSRD factors increase for the nearest W--O, Zn--O, Ni--O, W--Zn, W--Ni, Zn--Zn, and Ni--Ni interatomic pairs (see Fig.\ \ref{fig3}). In general, with the increasing temperature, no significant changes of the average interatomic distances were observed (see RDFs in insets in Fig.\ \ref{fig3}). However, the widths of RDFs -- characterised by the MSRD factors -- increase with temperature.

In both tungstates, \ce{[WO6]} octahedra are distorted due to the second-order Jahn-Teller effect \cite{KUNZ1995}, so that three groups of the W--O bonds can be distinguished \cite{Keeling1957, Filipenko1968, Trots2009}. The temperature-induced effects in \ce{[WO6]} octahedra are more pronounced in \ce{ZnWO4}: at low temperatures, three groups of W--O bonds can be identified at 1.8~\AA, 1.9~\AA\ and 2.1~\AA, whereas at high temperatures, the first two groups merge together and form a single wide peak in the RDFs $g_\text{W-O}(r)$ (see the inset in Fig.\ \ref{fig3}(b)). On the contrary, in \ce{NiWO4} (see the inset in Fig.\ \ref{fig3}(a)), the first two groups of W--O bonds produce a single peak in the range of 1.7--2.0~\AA\ with a weak temperature dependence. Note that W--O bonds are stiffer in \ce{ZnWO4}, as is evidenced by the larger value of the effective bond-stretching force constants (Table~\ref{tab1}) and of the larger frequency of the W--O stretching $\mathrm{A_g}$ mode position determined by Raman spectroscopy in \cite{Bakradze2020}. At the same time, the distant group composed of two oxygen atoms at around 2.1~\AA\ is significantly affected by thermal disorder both in \ce{ZnWO4} and \ce{NiWO4}, as reflected by the low values of the effective bond-stretching force constants (Table~\ref{tab1}).

\ce{[MO6]} (M = Ni, Zn) octahedra are less distorted than \ce{[WO6]} octahedra. The temperature dependence of the MSRD values for the M--O bonds in \ce{[MO6]} octahedra is very similar in \ce{ZnWO4} and \ce{NiWO4} (Fig.\ \ref{fig3}(c)), whereas the static contribution to the MSRD values for M--O bonds is larger in \ce{ZnWO4}. In the RDFs $g_\text{Zn-O}(r)$, a broader distribution of Zn--O bond lengths is present and it becomes more asymmetric at higher temperatures. On the contrary, \ce{[NiO6]} octahedra are almost regular with slightly different values of Ni--O bonds \cite{Keeling1957, Weitzel1976}. The difference in the temperature-dependence of the MSRDs for Zn--O and Ni--O bonds is small, however, the Zn--O bonds in \ce{ZnWO4} are less stronger -- i.e. have a smaller value of the effective bond-stretching force constant -- than Ni--O bonds in \ce{NiWO4} (Table\ \ref{tab1}).

A notable result was observed in the second coordination shell of nickel atoms (Fig.\ \ref{fig3}(d)): the RDFs $g_\text{Ni-Ni}(r)$ in \ce{NiWO4} exhibit two distinct maxima at all temperatures, indicating that -- as opposed to the average structure probed in diffraction experiments \cite{Keeling1957, Weitzel1976} -- there are two slightly different Ni--Ni interatomic distances with lengths equal to about 2.93~\AA\ and 3.05~\AA\ (see the insets in Fig.\ \ref{fig3}(d)). Note that according to the diffraction data, one would expect the closest cation-to-cation distances for all M--M cation pairs to be equal and slightly longer: around 3.06~\AA\ and 3.25~\AA\ for the Ni--Ni pairs in \ce{NiWO4} \cite{Keeling1957, Weitzel1976} and the Zn--Zn pairs in \ce{ZnWO4} \cite{Filipenko1968}, respectively (see the gray dashed vertical lines in the RDFs graphs in Fig.\ \ref{fig3}(d)). 

According to the semi-empirical Goodenough-Kanamori-Anderson rules \cite{Goodenough1976}, the magnetic coupling between nearest Ni$^{2+}$ (3d$^8$) ions in \ce{NiWO4} occurs through 90$\degree$ Ni$^{2+}$--O$^{2-}$--Ni$^{2+}$ superexchange interactions, in which the spins at nickel ions are ferromagnetically aligned within each chain of the \ce{[NiO6]} octahedra and the spins in neighbouring chains are ordered antiferromagnetically \cite{Weitzel1970, Weitzel1976}. A recent experimental study of lattice and spin dynamics in \ce{NiWO4} single-crystals using polarized Raman spectroscopy has shown that the local magnetic interactions in pure \ce{NiWO4} are determined by the spins at \ce{Ni^{2+}}(3d$^8$) ions through the superexchange paths involving \ce{[NiO6]} octahedra -- with Ni--O--Ni distances $\sim$2.06~\AA\ and the angle of about 95.7$\degree$ \cite{Prosnikov2017}. In the same work \cite{Prosnikov2017}, the magnetic excitation with the frequency of 24~cm$^{-1}$ has been found and tentatively assigned either to the Haldane gap \cite{Haldane1983} or spin dimer excitation \cite{Lake1996,Lake1997}.

The analysis of the structural configurations obtained from the RMC simulations indicates that the shorter and longer Ni--Ni distances alternate within the chains of \ce{[NiO6]} octahedra running along the [001]-directions. The local deviations from the average crystallographic structure are caused by the simultaneous displacements of nickel ions within \ce{[NiO6]} chains along the [010]-direction by $\sim$0.12~\AA\ and along the [001]-direction by about 0.02~\AA. The first displacement changes the nickel fractional coordinate $y$, which is a free parameter in the wolframite structure \cite{Keeling1957,Weitzel1976} so that the crystal symmetry is not affected. The former displacement induces an alternation of the Ni--Ni bond lengths, which become equal to 2.88~\AA\ and 2.97~\AA: the slightly larger values reported above can be explained by the thermal disorder effect, i.e. atom vibrations perpendicular to the interatomic bond \cite{Dalba1997,Fornasini2001}. However, the static displacement of nickel ions along the [001]-direction would result in the lowering of the crystal symmetry down to triclinic. Since it is not the case, we propose that vibration of nickel ions along the [001]-direction occurs in the double-well potential with the average nickel position coinciding with its fractional coordinate $z$(Ni)=0.25, observed by diffraction \cite{Keeling1957,Weitzel1976}. The MSRD $\sigma^2$ for the nearest nickel ions is almost temperature independent, suggesting a strong interaction (pairing) between them to exist at least up to 300 K (Fig.\ \ref{fig3}(d)). 

The magnetic ion pairing is known to exist in \ce{CuWO4} \cite{Lake1996,Lake1997,Koo2001} and was recently observed in the RDF $g_\text{Cu-Cu}(r)$\ as a splitting of the peak at about 3~\AA, determined from the RMC analysis of the Cu K-edge EXAFS (see \cite{timoshenko2014cuwo4} for details). \ce{CuWO4} crystalizes in triclinic ($P\bar{1}$) crystal structure, while other tungstates with small cations adopt monoclinic ($P2/c$) crystal structure. Magnetic Cu$^{2+}$(3d$^9$) cations octahedrally coordinated by oxygens experience the Jahn-Teller distortion, giving rise to a considerable lengthening of two opposite Cu--O bonds \cite{Ruiz2011,Kuzmin2013}, and, thus, a more distorted lattice of \ce{CuWO4} \cite{Kihlborg1970}. In \ce{CuWO4} structure, edge-sharing \ce{[CuO6]} octahedra form zigzag chains, which run along the [001]-direction and are responsible for low-dimensional magnetic properties of the tungstate. The axial distortion of \ce{[CuO6]} octahedra results in alternating shorter (2.99~\AA) and longer (3.14~\AA) Cu--Cu interatomic distances \cite{Forsyth1991}. As a result, two different superexchange interactions via the intrachain Cu--O--Cu bridges exist in \ce{CuWO4}, and the spin dimers occur between two nearest copper ions \cite{Lake1996,Lake1997,Koo2001}. Thus, a strong correlation exists in \ce{CuWO4} between its crystallographic and magnetic structures. At the same time, no Co--Co ion-pairing occurs in \ce{CoWO4} \cite{Weitzel1977}: according to the RMC analysis of the Co K-edge EXAFS, a single peak at about 3.1~\AA\ exists in the RDF $g_\text{Co-Co}(r)$ (see \cite{timoshenko2016cowo4} for details). In diamagnetic \ce{ZnWO4}, a single peak at about 3.23\AA\ was also observed in the RDFs $g_\text{Zn-Zn}(r)$\ (Fig.\ \ref{fig3}(d)) in agreement with the XRD data \cite{Trots2009}. The interaction between zinc ions is weaker than between nickel ones as is evidenced by the larger temperature dependence of its MSRD. 

Note that the RDFs $g_\text{Ni-Ni}(r)$ and $g_\text{Zn-Zn}(r)$ overlap in $R$-space with $g_\text{W-W}(r)$, and their reliable separation is possible within the RMC approach. Close values of the interatomic distances W--W result in a single peak in the RDFs at about 3.2~\AA\ (Fig.\ \ref{fig3}(f)). The W--W interactions are softer in \ce{ZnWO4} than in \ce{NiWO4} as evidenced from the temperature dependences of their MSRDs. The appearance of nickel-ion-pairing in \ce{NiWO4} could also influence the RDFs involving nickel and closely located tungsten ions (Fig.\ \ref{fig3}(e)). However, because the nickel-ion displacements are small the thermal disorder masks the pairing effect in the RDFs $g_\text{W-Ni}(r)$. According to the XRD data \cite{Keeling1957, Weitzel1976}, there are three groups of the interatomic distances W--Ni equal to 4$\times$3.48~\AA, 2$\times$3.57~\AA\ and 2$\times$3.72~\AA. The peaks in the RDFs $g_\text{W-Ni}(r)$\ are located at 6$\times$3.50~\AA\ and 2$\times$3.74~\AA\ with the first peak comprising two contributions. For comparison, the interatomic distances $R$(W--Zn) in \ce{ZnWO4} are equal to 2$\times$3.47~\AA, 2$\times$3.51~\AA, 2$\times$3.70~\AA\ and 2$\times$3.71~\AA\ \cite{Trots2009}. They contribute in pairs to two peaks in the RDFs with close amplitudes at low temperatures $g_\text{W-Zn}(r)$. A comparison of the MSRDs for W--Ni and W--Zn atom pairs indicates softer interactions between tungsten and zinc ions located at the distance of about 3.7~\AA. 

\subsection{Composition-induced effects}\label{ss:comp} 
EXAFS spectra of \ce{Zn_cNi_{1-c}WO4} solid solutions were measured at 300~K. The partial RDFs for different atomic pairs in \ce{Zn_cNi_{1-c}WO4} solid solutions were extracted from the final atomic configurations simulated by RMC and are presented in Fig.\ \ref{fig4}. As one can see, the formation of solid solutions, i.e. Ni substitution with Zn, has a pronounced effect both in the first and second coordination shells of metal ions.  

The shape of the RDFs $g_\text{Ni-O}(r)$ changes little for $c = 0.0-0.7$ (Fig.\ \ref{fig4}(a)): for $c=0.9$ the broad peak in the RDF $g_\text{Ni-O}(r)$ becomes split into two narrow components located at around 2.03~\AA\ and 2.18~\AA\ due to insufficient statistics caused by the low concentration of Ni atoms (indeed, there are only 12 Ni atoms in a $4 \times 4 \times 4$ simulation cell with a total of 768 atoms).

The RDFs $g_\text{Zn-O}(r)$ are broader than $g_\text{Ni-O}(r)$ but also demonstrate a weak composition dependence as seen in Fig.\ \ref{fig4}(b). A distinct shoulder at around 2.3~\AA\ is present in $g_\text{Zn-O}(r)$ of solid solutions but it becomes less pronounced in the pure \ce{ZnWO4}. This fact indicates the structural role of nickel ions, which can organize their local environment, while zinc ions simply adapt to it.  

The RDFs $g_\text{W-O}(r)$ show some composition dependence and consist of two well-separated peaks at about 1.80~\AA\ and 2.15~\AA\ composed of four and two oxygen atoms, respectively (Fig.\ \ref{fig4}(b)). The behaviour of the nearest group of oxygen atoms around tungsten upon Ni substitution with Zn correlates with that of the W--O stretching mode $\mathrm{A_g}$ at about 900~cm$^{-1}$ in the Raman scattering spectra \cite{Bakradze2020}. An increase of the W--O stretching mode frequency from 890~cm$^{-1}$ in \ce{NiWO4} to 907~cm$^{-1}$ in \ce{ZnWO4} \cite{Bakradze2020} suggests the presence of stronger (and shorter) bonds between W and nearest O atoms in the Zn-rich samples, which manifests itself in the RDFs as a small shift of the first peak maximum from 1.84~\AA\ to 1.80~\AA\ (whereas the bonds to the distant O atoms become slightly longer, i.e. less strongly bound) (Fig.\ \ref{fig4}(c)). This interpretation is backed up by the results of the MSRD analysis in terms of the characteristic Einstein frequencies and bond-stretching constants (Table\ \ref{tab1}), indicating that  the W--O bonds are stiffer in pure \ce{ZnWO4} than in \ce{NiWO4}. At the same time, an increase of the W--O stretching mode frequency for samples with a higher Zn content indicates a weakening of more covalent Ni--O bonds: Ni--O bonds exhibit a certain degree of covalency because \ce{Ni^{2+}} ions (\ce{[Ar]{3}d^{8}{4}s^2}) with spin $S = +1$ have a half-filled $\mathrm{e_g}$ orbital. 

The RDFs for Ni(Zn,W)--Ni(Zn) atom pairs demonstrate strong composition dependence since the RDF-peak area is equal to the respective coordination numbers, which change upon substitution of Ni by Zn atoms. The peak positions in the RDFs $g_\text{W-Ni}(r)$, $g_\text{W-Zn}(r)$, and $g_\text{Zn-Zn}(r)$ agree with the expected crystallographic values, indicated in Fig.\ \ref{fig4} by the vertical dashed lines. The most interesting behaviour is observed in the composition dependence of the RDFs $g_\text{Ni-Ni}(r)$. In pure \ce{NiWO4} ($c=0$), it has a double-peak shape (see also the inset in Fig.\ \ref{fig3}(d)), which survives until $c=0.6$, while at high Zn content only a single peak remains as in the RDF $g_\text{Zn-Zn)}(r)$ for pure \ce{ZnWO4}. The origin of two different Ni--Ni distances is related to the local magnetic interactions determined by the spins of \ce{Ni^{2+}} ions through the super-exchange paths involving the edge-sharing \ce{[NiO6]} octahedra in zigzag chains running along the [001]-direction. A probability that a Ni ion will have $n$ Ni ions as nearest neighbours in solid solution is given by the Poisson distribution $\displaystyle N! \frac{(1-c)^n c^{N-n}}{(N-n)! n!}$, where $N$ is the coordination number. The substitution occurs within quasi-one dimensional zigzag chains, therefore Ni ions can have only two nearest neighbours of Ni ions, i.e. $N$=2, and the probability to find two Ni atoms in neighbouring positions is given by $(1-c)^2$ and quickly vanishes with decreasing $c$ and the formation of Ni dimers is not likely already at $c=0.6$ (Fig.\ \ref{fig4}(d)).

\section{Conclusions}\label{s:conc}
Advanced analysis of the EXAFS spectra of three metal absorption edges simultaneously using the reverse Monte Carlo method and confined to a single structural model allowed us to monitor the influence of thermal disorder and static distortions on the local structure of microcrystalline \ce{Zn_cNi_{1-c}WO4} ($c=0.0-1.0$) solid solutions. 

In \ce{ZnWO4}, at low temperatures, three groups of W--O bonds can be identified with the bond lengths being equal to 1.8~\AA, 1.9~\AA, and 2.15~\AA\; whereas, in \ce{NiWO4}, two groups of W--O bonds can be identified with the bond lengths being equal to 1.8~\AA\ and 2.15~\AA. Thus, the distortion of \ce{[WO6]} octahedra is more pronounced in \ce{ZnWO4} than in \ce{NiWO4}. In both pure tungstates, thermal effects mainly influence the distant group of oxygen atoms located at $R$(W--O)=2.15~\AA. While both Zn and Ni are octahedrally coordinated by oxygen atoms, the first coordination shell of zinc atoms is more distorted as one can conclude from a more asymmetric shape of the RDF $g_\text{Zn-O}(r)$. 

The composition-induced distortions of \ce{[WO6]}, \ce{[NiO6]} and \ce{[ZnO6]} octahedra in \ce{Zn_cNi_{1-c}WO4} solid solutions at 300~K were evidenced from the shape of the W--O, Ni--O, and Zn--O radial distribution functions, respectively. In \ce{Zn_cNi_{1-c}WO4} solid solutions, the Zn concentration has a pronounced effect on the polyhedral units. The observed changes are caused by the subtle interplay between Ni--O, Zn--O, and W--O interaction. With increasing Zn content, the lengths of the nearest W--O bonds become shorter, while the lengths of the distant W--O bonds become slightly longer. \ce{[NiO6]} octahedra almost insensitive to the addition of Zn atoms, whereas \ce{[ZnO6]} octahedra show a weak composition dependence and are more distorted in solid solutions than in pure \ce{ZnWO4}. 

The most interesting result has been found in the second coordination shell of nickel atoms. In pure \ce{NiWO4}, dimerization of \ce{Ni^{2+}} ions within quasi-one-dimensional zigzag chains running along the $c$-axis directions was observed in the whole studied temperature range. It results in the two slightly different Ni--Ni interatomic distances with lengths equal to about 2.9~\AA\ and 3.0~\AA. The existence of \ce{Ni^{2+}} ion dimers was speculated recently in \cite{Prosnikov2017} to explain the magnetic origin of the low frequency (24~cm$^{-1}$) band in the Raman scattering spectra of \ce{NiWO4}. The observed dimerization of nickel ions in pure \ce{NiWO4} is preserved in solid solutions \ce{Zn_cNi_{1-c}WO_4} for $c \leq 0.6$, which is related to the probability to find two Ni$^{2+}$ ions in neighbouring positions. 

To conclude, we have demonstrated on the example of pure tungstates and their solid solutions that an accurate analysis of EXAFS data by the reverse Monte Carlo method allows one to probe  distant interatomic interactions in the region of the medium-range order leading to the local structural deviations from the average crystallographic structure.

\section*{Declaration of Competing Interest}

The authors declare that they have no known competing financial interests or personal relationships that could have appeared to influence the work reported in this paper.

\section*{Acknowledgements}
G.B. acknowledges the financial support provided by the State Education Development Agency for project No. 1.1.1.2/VIAA/3/19/444 (agreement No. 1.1.1.2/16/I/001) realized at the Institute of Solid State Physics, University of Latvia. 
A.K. and A.K. would like to thank the support of the Latvian Council of Science project No. lzp-2019/1-0071. 
Institute of Solid State Physics, University of Latvia as the Center of Excellence has received funding from the European Union’s Horizon 2020 Framework Programme H2020-WIDESPREAD-01-2016-2017-TeamingPhase2 under grant agreement No. 739508, project CAMART2.

\newpage

%\bibliography{ZnNiWO4}

\begin{thebibliography}{10}
	\expandafter\ifx\csname url\endcsname\relax
	\def\url#1{\texttt{#1}}\fi
	\expandafter\ifx\csname urlprefix\endcsname\relax\def\urlprefix{URL }\fi
	\expandafter\ifx\csname href\endcsname\relax
	\def\href#1#2{#2} \def\path#1{#1}\fi
	
	\bibitem{Pettifer2005}
	R.~Pettifer, O.~Mathon, S.~Pascarelli, M.~D. Cooke, M.~R.~J. Gibbs, Measurement
	of femtometre-scale atomic displacements by x-ray absorption spectroscopy,
	Nature 435 (2005) 78--81.
	\newblock \href {https://doi.org/10.1038/nature03516}
	{\path{doi:10.1038/nature03516}}.
	
	\bibitem{Korobko2015}
	R.~Korobko, A.~Lerner, Y.~Li, E.~Wachtel, A.~I. Frenkel, I.~Lubomirsky, In-situ
	extended x-ray absorption fine structure study of electrostriction in gd
	doped ceria, Appl. Phys. Lett. 106 (2015) 042904.
	\newblock \href {https://doi.org/10.1063/1.4906857}
	{\path{doi:10.1063/1.4906857}}.
	
	\bibitem{Purans2008ge}
	J.~Purans, N.~D. Afify, G.~Dalba, R.~Grisenti, S.~De~Panfilis, A.~Kuzmin, V.~I.
	Ozhogin, F.~Rocca, A.~Sanson, S.~I. Tiutiunnikov, P.~Fornasini, Isotopic
	effect in extended x-ray-absorption fine structure of germanium, Phys. Rev.
	Lett. 100 (2008) 055901.
	\newblock \href {https://doi.org/10.1103/PhysRevLett.100.055901}
	{\path{doi:10.1103/PhysRevLett.100.055901}}.
	
	\bibitem{Nemeth2012}
	K.~N\'emeth, K.~W. Chapman, M.~Balasubramanian, B.~Shyam, P.~J. Chupas, S.~M.
	Heald, M.~Newville, R.~J. Klingler, R.~E. Winans, J.~D. Almer, G.~Sandi,
	G.~Srajer, {Efficient simultaneous reverse Monte Carlo modeling of
		pair-distribution functions and extended x-ray-absorption fine structure
		spectra of crystalline disordered materials}, J. Chem. Phys. 136 (2012)
	074105.
	\newblock \href {https://doi.org/10.1063/1.3684547}
	{\path{doi:10.1063/1.3684547}}.
	
	\bibitem{Timoshenko2012rmc}
	J.~Timoshenko, A.~Kuzmin, J.~Purans, {Reverse Monte Carlo modeling of thermal
		disorder in crystalline materials from EXAFS spectra}, Comp. Phys. Commun.
	183 (2012) 1237--1245.
	\newblock \href {https://doi.org/10.1016/j.cpc.2012.02.002}
	{\path{doi:10.1016/j.cpc.2012.02.002}}.
	
	\bibitem{Timoshenko2014zno}
	J.~Timoshenko, A.~Anspoks, A.~Kalinko, A.~Kuzmin, {Temperature dependence of
		the local structure and lattice dynamics of wurtzite-type ZnO}, Acta Mater.
	79 (2014) 194--202.
	\newblock \href {https://doi.org/10.1016/j.actamat.2014.07.029}
	{\path{doi:10.1016/j.actamat.2014.07.029}}.
	
	\bibitem{Timoshenko2017a}
	J.~Timoshenko, A.~I. Frenkel, {Probing structural relaxation in nanosized
		catalysts by combining EXAFS and reverse Monte Carlo methods}, Catal. Today
	280 (2017) 274--282.
	\newblock \href {https://doi.org/10.1016/j.cattod.2016.05.049}
	{\path{doi:10.1016/j.cattod.2016.05.049}}.
	
	\bibitem{Timoshenko2017c}
	J.~Timoshenko, A.~Anspoks, A.~Kalinko, A.~Kuzmin, {Thermal disorder and
		correlation effects in anti-perovskite-type copper nitride}, Acta Mater. 129
	(2017) 61--71.
	\newblock \href {https://doi.org/10.1016/j.actamat.2017.02.074}
	{\path{doi:10.1016/j.actamat.2017.02.074}}.
	
	\bibitem{Levin2017}
	I.~Levin, V.~Krayzman, G.~Cibin, M.~G. Tucker, M.~Eremenko, K.~Chapman, R.~L.
	Paul, {Coupling of emergent octahedral rotations to polarization in
		(K,Na)NbO$_3$ ferroelectrics}, Sci. Rep. 7 (2017) 15620.
	\newblock \href {https://doi.org/10.1038/s41598-017-15937-x}
	{\path{doi:10.1038/s41598-017-15937-x}}.
	
	\bibitem{DiCicco2018}
	A.~{Di Cicco}, F.~Iesari, A.~Trapananti, P.~{D’Angelo}, A.~Filipponi,
	{Structure and atomic correlations in molecular systems probed by XAS reverse
		Monte Carlo refinement}, J. Chem. Phys. 148 (2018) 094307.
	\newblock \href {https://doi.org/10.1063/1.5013660}
	{\path{doi:10.1063/1.5013660}}.
	
	\bibitem{Kraynis2019}
	O.~Kraynis, J.~Timoshenko, J.~Huang, H.~Singh, E.~Wachtel, A.~I. Frenkel,
	I.~Lubomirsky, Modeling strain distribution at the atomic level in doped
	ceria films with extended x-ray absorption fine structure spectroscopy,
	Inorg. Chem. 58 (2019) 7527--7536.
	\newblock \href {https://doi.org/10.1021/acs.inorgchem.9b00730}
	{\path{doi:10.1021/acs.inorgchem.9b00730}}.
	
	\bibitem{Pethes2019}
	I.~Pethes, V.~Nazabal, J.~Ari, I.~Kaban, J.~Darpentigny, E.~Welter,
	O.~Gutowski, B.~Bureau, Y.~Messaddeq, P.~J\'ov\'ari, {Atomic level structure
		of Ge-Sb-S glasses: Chemical short range order and long Sb-S bonds}, J.
	Alloys Compd. 774 (2019) 1009--1016.
	\newblock \href {https://doi.org/10.1016/j.jallcom.2018.09.334}
	{\path{doi:10.1016/j.jallcom.2018.09.334}}.
	
	\bibitem{Rehr2000}
	J.~J. Rehr, R.~C. Albers, Theoretical approaches to x-ray absorption fine
	structure, Rev. Mod. Phys. 72 (2000) 621--654.
	\newblock \href {https://doi.org/10.1103/RevModPhys.72.621}
	{\path{doi:10.1103/RevModPhys.72.621}}.
	
	\bibitem{Natoli2003}
	C.~R. Natoli, M.~Benfatto, S.~Della~Longa, K.~Hatada, {X-ray absorption
		spectroscopy: state-of-the-art analysis}, J. Synchrotron Rad. 10 (2003)
	26--42.
	\newblock \href {https://doi.org/10.1107/S0909049502017247}
	{\path{doi:10.1107/S0909049502017247}}.
	
	\bibitem{Rehr2009}
	J.~J. Rehr, J.~J. Kas, M.~P. Prange, A.~P. Sorini, Y.~Takimoto, F.~Vila, {Ab
		initio theory and calculations of X-ray spectra}, C. R. Phys. 10 (2009)
	548--559.
	\newblock \href {https://doi.org/10.1016/j.crhy.2008.08.004}
	{\path{doi:10.1016/j.crhy.2008.08.004}}.
	
	\bibitem{timoshenko2014cuwo4}
	J.~Timoshenko, A.~Anspoks, A.~Kalinko, A.~Kuzmin, {Analysis of extended x-ray
		absorption fine structure data from copper tungstate by the reverse Monte
		Carlo method}, Phys. Scr. 89 (2014) 044006.
	\newblock \href {https://doi.org/10.1088/0031-8949/89/04/044006}
	{\path{doi:10.1088/0031-8949/89/04/044006}}.
	
	\bibitem{timoshenko2015mncowo4}
	J.~Timoshenko, A.~Anspoks, A.~Kalinko, I.~Jonane, A.~Kuzmin, {Local structure
		of multiferroic MnWO$_4$ and Mn$_{0.7}$Co$_{0.3}$WO$_4$ revealed by the
		evolutionary algorithm}, Ferroelectrics 483 (2015) 68--74.
	\newblock \href {https://doi.org/10.3204/PUBDB-2015-04735}
	{\path{doi:10.3204/PUBDB-2015-04735}}.
	
	\bibitem{timoshenko2016cowo4}
	J.~Timoshenko, A.~Anspoks, A.~Kalinko, A.~Kuzmin, {Local structure of cobalt
		tungstate revealed by EXAFS spectroscopy and reverse Monte Carlo/evolutionary
		algorithm simulations}, Z. Phys. Chem. 230 (2016) 551--568.
	\newblock \href {https://doi.org/10.1515/zpch-2015-0646}
	{\path{doi:10.1515/zpch-2015-0646}}.
	
	\bibitem{Keeling1957}
	R.~O. Keeling~Jr., {The structure of NiWO$_4$}, Acta Crystallogr. 10 (1957)
	209--213.
	\newblock \href {https://doi.org/10.1107/s0365110x57000651}
	{\path{doi:10.1107/s0365110x57000651}}.
	
	\bibitem{Filipenko1968}
	O.~S. Filipenko, E.~A. Pobedimskaya, N.~V. Belov, {Crystal structure of
		ZnWO$_4$}, Sov. Phys. Crystallography 13 (1968) 127--129.
	
	\bibitem{Trots2009}
	D.~Trots, A.~Senyshyn, L.~Vasylechko, R.~Dr, T.~Vad, V.~Mikhailik, H.~Kraus,
	{Crystal structure of ZnWO$_4$ scintillator material in the range of 3-1423
		K}, J. Phys.: Condens. Matter 21 (2009) 325402.
	\newblock \href {https://doi.org/10.1088/0953-8984/21/32/325402}
	{\path{doi:10.1088/0953-8984/21/32/325402}}.
	
	\bibitem{KUNZ1995}
	M.~Kunz, I.~Brown, {Out-of-Center Distortions around Octahedrally Coordinated
		d$^0$ Transition Metals}, J. Solid State Chem. 115 (1995) 395--406.
	\newblock \href {https://doi.org/10.1006/jssc.1995.1150}
	{\path{doi:10.1006/jssc.1995.1150}}.
	
	\bibitem{Oliveira2008}
	A.~L.~M. de~Oliveira, J.~M. Ferreira, M.~R.~S. Silva, G.~S. Braga, L.~E.~B.
	Soledade, M.~A. M.~M. Aldeiza, C.~A. Paskocimas, S.~J.~G. Lima, E.~Longo,
	A.~G. de~Souza, I.~M.~G. dos Santos, {Yellow Zn$_{x}$Ni$_{1-x}$WO$_4$
		pigments obtained using a polymeric precursor method}, Dyes Pigm. 77 (2008)
	210--216.
	\newblock \href {https://doi.org/10.1016/j.dyepig.2007.05.004}
	{\path{doi:10.1016/j.dyepig.2007.05.004}}.
	
	\bibitem{Liu1988}
	Y.~Liu, H.~Wang, G.~Chen, Y.~D. Zhou, B.~Y. Gu, B.~Q. Hu, {Analysis of Raman
		spectra of ZnWO$_4$ single crystals}, J. Appl. Phys. 64 (1988) 4651--4653.
	\newblock \href {https://doi.org/10.1063/1.341245}
	{\path{doi:10.1063/1.341245}}.
	
	\bibitem{Wang1992}
	H.~Wang, F.~D. Medina, Y.~D. Zhou, Q.~N. Zhang, {Temperature dependence of the
		polarized Raman spectra of ZnWO$_4$ single crystals}, Phys. Rev. B 45 (1992)
	10356--10362.
	\newblock \href {https://doi.org/10.1103/PhysRevB.45.10356}
	{\path{doi:10.1103/PhysRevB.45.10356}}.
	
	\bibitem{Fomichev1994}
	V.~Fomichev, O.~Kondratov, {Vibrational spectra of compounds with the
		wolframite structure}, Spectrochim. Acta A 50 (1994) 1113--1120.
	\newblock \href {https://doi.org/10.1016/0584-8539(94)80034-0}
	{\path{doi:10.1016/0584-8539(94)80034-0}}.
	
	\bibitem{Kuzmin2001a}
	A.~Kuzmin, J.~Purans, R.~Kalendarev, {Local structure and vibrational dynamics
		in NiWO$_4$}, Ferroelectrics 258 (2001) 21--30.
	\newblock \href {https://doi.org/10.1080/00150190108008653}
	{\path{doi:10.1080/00150190108008653}}.
	
	\bibitem{Errandonea2008}
	D.~Errandonea, F.~J. Manj\'on, N.~Garro, P.~Rodr\'{\i}guez-Hern\'andez,
	S.~Radescu, A.~Mujica, A.~Mu\~noz, C.~Y. Tu, {Combined Raman scattering and
		ab initio investigation of pressure-induced structural phase transitions in
		the scintillator ZnWO$_4$}, Phys. Rev. B 78 (2008) 054116.
	\newblock \href {https://doi.org/10.1103/PhysRevB.78.054116}
	{\path{doi:10.1103/PhysRevB.78.054116}}.
	
	\bibitem{Kuzmin2011niwo4}
	A.~Kuzmin, A.~Kalinko, R.~A. Evarestov, {First-principles LCAO study of phonons
		in NiWO$_4$}, Centr. Eur. J. Phys. 9 (2011) 502--509.
	\newblock \href {https://doi.org/10.2478/s11534-010-0091-z}
	{\path{doi:10.2478/s11534-010-0091-z}}.
	
	\bibitem{Wilkinson1977}
	C.~Wilkinson, M.~J. Sprague, {The magnetic structures of NiWO$_4$ and
		CoWO$_4$}, Z. Kristallogr. 145 (1977) 96--107.
	\newblock \href {https://doi.org/10.1524/zkri.1977.145.1-2.96}
	{\path{doi:10.1524/zkri.1977.145.1-2.96}}.
	
	\bibitem{Prosnikov2017}
	M.~A. Prosnikov, V.~Y. Davydov, A.~N. Smirnov, M.~P. Volkov, R.~V. Pisarev,
	P.~Becker, L.~Bohat\'y, {Lattice and spin dynamics in a low-symmetry
		antiferromagnet ${\mathrm{NiWO}}_{4}$}, Phys. Rev. B 96 (2017) 014428.
	\newblock \href {https://doi.org/10.1103/PhysRevB.96.014428}
	{\path{doi:10.1103/PhysRevB.96.014428}}.
	
	\bibitem{Liu2017}
	C.~Liu, Z.~He, Y.~Liu, R.~Chen, M.~Shi, H.~Zhu, C.~Dong, J.~Wang, {Magnetic
		anisotropy and spin-flop transition of NiWO$_4$ single crystals}, J. Magn.
	Magn. Mater. 444 (2017) 190--192.
	\newblock \href {https://doi.org/10.1016/j.jmmm.2017.08.032}
	{\path{doi:10.1016/j.jmmm.2017.08.032}}.
	
	\bibitem{Weitzel1970}
	von Hans~Weitzel, {Magnetische struktur von CoWO$_4$, NiWO$_4$ und CuWO$_4$},
	Solid State Commun. 8 (1970) 2071--2072.
	\newblock \href {https://doi.org/10.1016/0038-1098(70)90221-8}
	{\path{doi:10.1016/0038-1098(70)90221-8}}.
	
	\bibitem{Weitzel1976}
	von Hans~Weitzel, {Kristallstrukturverfeinerung von Wolframiten und
		Columbiten}, Z. Kristallogr. 144 (1976) 238--258.
	\newblock \href {https://doi.org/10.1524/zkri.1976.144.16.238}
	{\path{doi:10.1524/zkri.1976.144.16.238}}.
	
	\bibitem{Shannon1976}
	R.~D. Shannon, Revised effective ionic radii and systematic studies of
	interatomic distances in halides and chalcogenides, Acta Cryst. A 32 (1976)
	751--767.
	\newblock \href {https://doi.org/10.1107/S0567739476001551}
	{\path{doi:10.1107/S0567739476001551}}.
	
	\bibitem{Karmakar2020}
	S.~Karmakar, D.~Behera, {High-temperature impedance and alternating current
		conduction mechanism of Ni$_{0.5}$Zn$_{0.5}$WO$_4$ micro-crystal for
		electrical energy storage application}, J. Aust. Ceram. Soc. 56 (2020)
	1253--1259.
	\newblock \href {https://doi.org/10.1007/s41779-020-00475-z}
	{\path{doi:10.1007/s41779-020-00475-z}}.
	
	\bibitem{Kalinko2011a}
	A.~Kalinko, A.~Kotlov, A.~Kuzmin, V.~Pankratov, A.~I. Popov, L.~Shirmane,
	{Electronic excitations in ZnWO$_4$ and Zn$_x$Ni$_{1-x}$WO$_4$ using VUV
		synchrotron radiation}, Centr. Eur. J. Phys. 9 (2011) 432--437.
	\newblock \href {https://doi.org/10.2478/s11534-010-0108-7}
	{\path{doi:10.2478/s11534-010-0108-7}}.
	
	\bibitem{Huang2015}
	H.~Huang, L.~Liu, N.~Tian, Y.~Zhang, {Structure, optical properties, and
		magnetism of Zn$_{1-x}$Ni$_x$WO$_4$ ($0 \leq x \leq 1$) solid solution}, J.
	Alloys Compd. 637 (2015) 471--475.
	\newblock \href {https://doi.org/10.1016/j.jallcom.2015.02.224}
	{\path{doi:10.1016/j.jallcom.2015.02.224}}.
	
	\bibitem{MnWO4}
	G.~Lautenschl\"ager, H.~Weitzel, T.~Vogt, R.~Hock, A.~B\"ohm, M.~Bonnet,
	H.~Fuess, {Magnetic phase transitions of MnWO$_4$ studied by the use of
		neutron diffraction}, Phys. Rev. B 48 (1993) 6087--6098.
	\newblock \href {https://doi.org/10.1103/PhysRevB.48.6087}
	{\path{doi:10.1103/PhysRevB.48.6087}}.
	
	\bibitem{Lake1996}
	B.~Lake, D.~A. Tennant, R.~A. Cowley, J.~D. Axe, C.~K. Chen, {Magnetic
		excitations in the ordered phase of the antiferromagnetic alternating chain
		compound}, J. Phys.: Condens. Matter 8 (1996) 8613--8634.
	\newblock \href {https://doi.org/10.1088/0953-8984/8/44/013}
	{\path{doi:10.1088/0953-8984/8/44/013}}.
	
	\bibitem{Lake1997}
	B.~Lake, R.~A. Cowley, D.~A. Tennant, {A dimer theory of the magnetic
		excitations in the ordered phase of the alternating-chain compound}, J.
	Phys.: Condens. Matter 9 (1997) 10951--10975.
	\newblock \href {https://doi.org/10.1088/0953-8984/9/49/014}
	{\path{doi:10.1088/0953-8984/9/49/014}}.
	
	\bibitem{Koo2001}
	H.-J. Koo, M.-H. Whangbo, {Spin dimer analysis of the anisotropic spin exchange
		interactions in the distorted wolframite-type oxides CuWO$_4$, CuMoO$_4$-III,
		and Cu(Mo$_{0.25}$W$_{0.75}$)O$_4$}, Inorg. Chem. 40 (2001) 2161--2169.
	\newblock \href {https://doi.org/10.1021/ic001445r}
	{\path{doi:10.1021/ic001445r}}.
	
	\bibitem{Bakradze2020}
	G.~Bakradze, A.~Kalinko, A.~Kuzmin, {X-ray absorption and Raman spectroscopy
		studies of tungstates solid solutions Zn cNi1- cWO4(c = 0.0-1.0)}, Low Temp.
	Phys. 46 (2020) 1201--1205.
	\newblock \href {https://doi.org/10.1063/10.0002474}
	{\path{doi:10.1063/10.0002474}}.
	
	\bibitem{DORISC}
	K.~Rickers, W.~Drube, H.~Schulte-Schrepping, E.~Welter, U.~Br\"{u}ggmann,
	M.~Herrmann, J.~Heuer, H.~Schulz-Ritter, {New XAFS facility for in-situ
		measurements at beamline C at HASYLAB}, AIP Conf. Proc. 882 (2007) 905.
	\newblock \href {https://doi.org/10.1063/1.2644700}
	{\path{doi:10.1063/1.2644700}}.
	
	\bibitem{Kuzmin2014}
	A.~Kuzmin, J.~Chaboy, {EXAFS and XANES analysis of oxides at the nanoscale},
	IUCrJ 1 (2014) 571--589.
	\newblock \href {https://doi.org/10.1107/S2052252514021101}
	{\path{doi:10.1107/S2052252514021101}}.
	
	\bibitem{Timoshenko2014rmc}
	J.~Timoshenko, A.~Kuzmin, J.~Purans, {EXAFS study of hydrogen intercalation
		into ReO$_3$ using the evolutionary algorithm}, J. Phys.: Condens. Matter 26
	(2014) 055401.
	\newblock \href {https://doi.org/10.1088/0953-8984/26/5/055401}
	{\path{doi:10.1088/0953-8984/26/5/055401}}.
	
	\bibitem{Ankudinov1998}
	A.~L. Ankudinov, B.~Ravel, J.~J. Rehr, S.~D. Conradson, {Real-space
		multiple-scattering calculation and interpretation of x-ray-absorption
		near-edge structure}, Phys. Rev. B 58 (1998) 7565--7576.
	\newblock \href {https://doi.org/10.1103/PhysRevB.58.7565}
	{\path{doi:10.1103/PhysRevB.58.7565}}.
	
	\bibitem{Timoshenko2009wavelet}
	J.~Timoshenko, A.~Kuzmin, {Wavelet data analysis of EXAFS spectra}, Comp. Phys.
	Commun. 180 (2009) 920--925.
	\newblock \href {https://doi.org/10.1016/j.cpc.2008.12.020}
	{\path{doi:10.1016/j.cpc.2008.12.020}}.
	
	\bibitem{Sevillano1979}
	E.~Sevillano, H.~Meuth, J.~Rehr, {Extended x-ray absorption fine structure
		Debye-Waller factors. I. Monatomic crystals}, Phys. Rev. B 20 (1979) 4908.
	\newblock \href {https://doi.org/10.1103/PhysRevB.20.4908}
	{\path{doi:10.1103/PhysRevB.20.4908}}.
	
	\bibitem{Goodenough1976}
	J.~B. Goodenough, {Magnetism and the Chemical Bond}, 1st Edition, Vol.~1 of
	Interscience Monographs on Chemistry, Inorganic Chemistry section, R. E.
	Krieger Pub. Co, 1976.
	
	\bibitem{Haldane1983}
	F.~Haldane, {Continuum dynamics of the 1-D Heisenberg antiferromagnet:
		Identification with the O(3) nonlinear sigma model}, Phys. Lett. A 93 (1983)
	464--468.
	\newblock \href {https://doi.org/10.1016/0375-9601(83)90631-X}
	{\path{doi:10.1016/0375-9601(83)90631-X}}.
	
	\bibitem{Dalba1997}
	G.~Dalba, P.~Fornasini, {EXAFS Debye-Waller factor and thermal vibrations of
		crystals}, J. Synchrotron Rad. 4 (1997) 243--255.
	\newblock \href {https://doi.org/10.1107/S0909049597006900}
	{\path{doi:10.1107/S0909049597006900}}.
	
	\bibitem{Fornasini2001}
	P.~Fornasini, {Study of lattice dynamics via extended X-ray absorption fine
		structure}, J. Phys.: Condensed Matter 13 (2001) 7859--7872.
	\newblock \href {https://doi.org/10.1088/0953-8984/13/34/324}
	{\path{doi:10.1088/0953-8984/13/34/324}}.
	
	\bibitem{Ruiz2011}
	J.~Ruiz-Fuertes, A.~Friedrich, J.~Pellicer-Porres, D.~Errandonea, A.~Segura,
	W.~Morgenroth, E.~Hauss\"uhl, C.-Y. Tu, A.~Polian, {Structure solution of the
		high-pressure phase of CuWO$_4$ and evolution of the Jahn-Teller distortion},
	Chem. Mater. 23 (2011) 4220--4226.
	\newblock \href {https://doi.org/10.1021/cm201592h}
	{\path{doi:10.1021/cm201592h}}.
	
	\bibitem{Kuzmin2013}
	A.~Kuzmin, A.~Kalinko, R.~Evarestov, {Ab initio LCAO study of the atomic,
		electronic and magnetic structures and the lattice dynamics of triclinic
		CuWO$_4$}, Acta Mater. 61 (2013) 371--378.
	\newblock \href {https://doi.org/10.1016/j.actamat.2012.10.002}
	{\path{doi:10.1016/j.actamat.2012.10.002}}.
	
	\bibitem{Kihlborg1970}
	L.~Kihlborg, E.~Gebert, {CuWO$_4$, a distorted Wolframite-type structure}, Acta
	Crystallogr. B 26 (1970) 1020--1026.
	\newblock \href {https://doi.org/10.1107/S0567740870003515}
	{\path{doi:10.1107/S0567740870003515}}.
	
	\bibitem{Forsyth1991}
	J.~B. Forsyth, C.~Wilkinson, A.~I. Zvyagin, {The antiferromagnetic structure of
		copper tungstate CuWO$_4$}, J. Phys.: Condens. Matter 3 (1991) 8433.
	\newblock \href {https://doi.org/10.1088/0953-8984/3/43/010}
	{\path{doi:10.1088/0953-8984/3/43/010}}.
	
	\bibitem{Weitzel1977}
	H.~Weitzel, H.~Langhof, {Refinement of the magnetic structures of NiWO$_4$ and
		CoWO$_4$}, J. Magn. Magn. Mater. 4 (1977) 265--274.
	\newblock \href {https://doi.org/10.1016/0304-8853(77)90047-6}
	{\path{doi:10.1016/0304-8853(77)90047-6}}.
	
\end{thebibliography}

\newpage

\appendix

\section{RMC/EA simulation results for pure \ce{NiWO4} and \ce{ZnWO4}}
\begin{figure*}[t]
	\centering
	\includegraphics[width=0.95\textwidth]{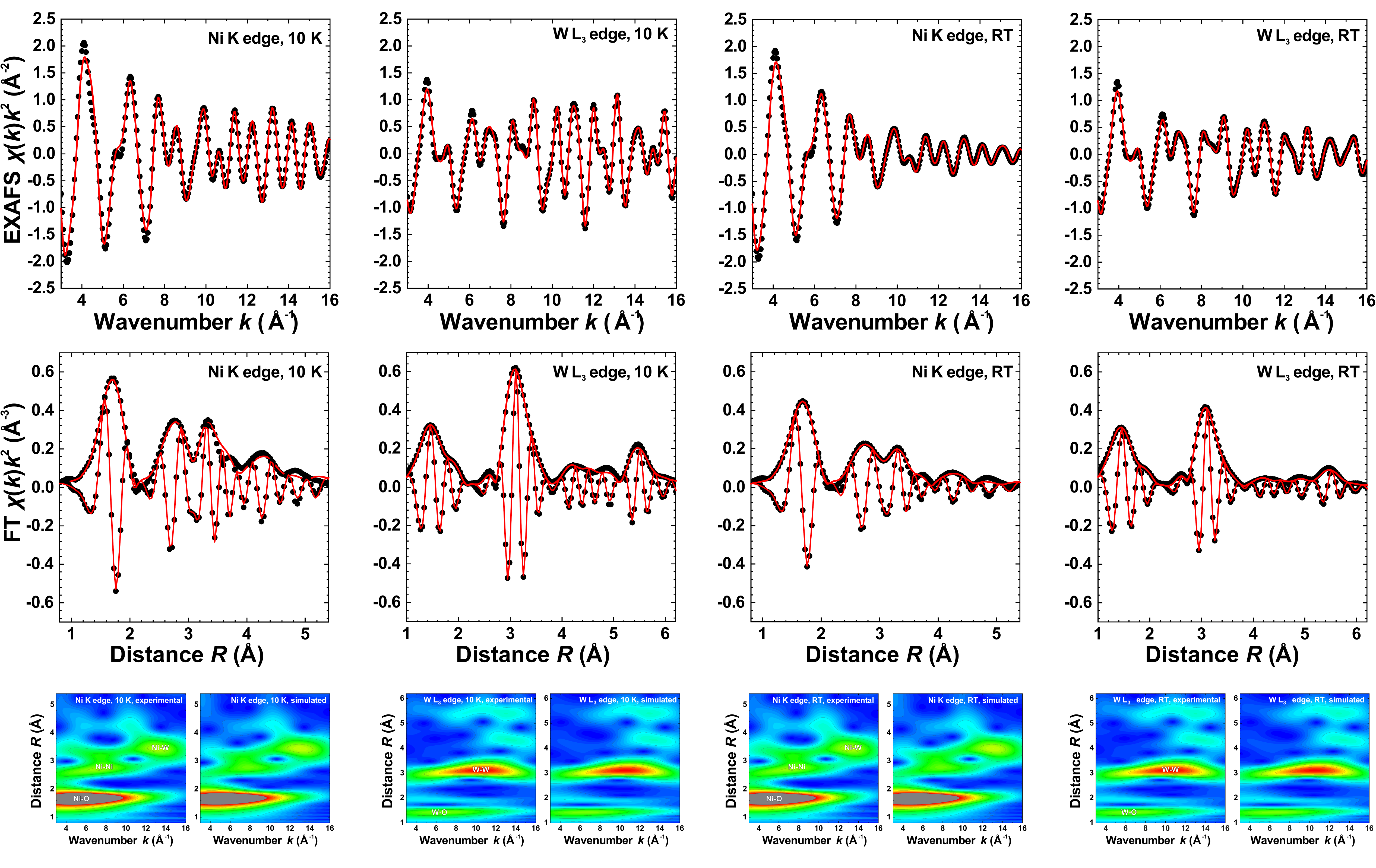}	
	\caption{Experimental (black dots) and RMC/EA calculated (solid lines) EXAFS spectra $\chi(k)k^2$ of pure \ce{NiWO4} at the Ni K-edges and W L$_3$-edge (upper row), their Fourier transforms (FTs) (middle row) and wavelet transforms (lower rows) at 10~K and 300~K. Both modulus and imaginary parts are shown in FTs.}
	\label{figa1}
\end{figure*}

\begin{figure*}
	\centering
	\includegraphics[width=0.95\textwidth]{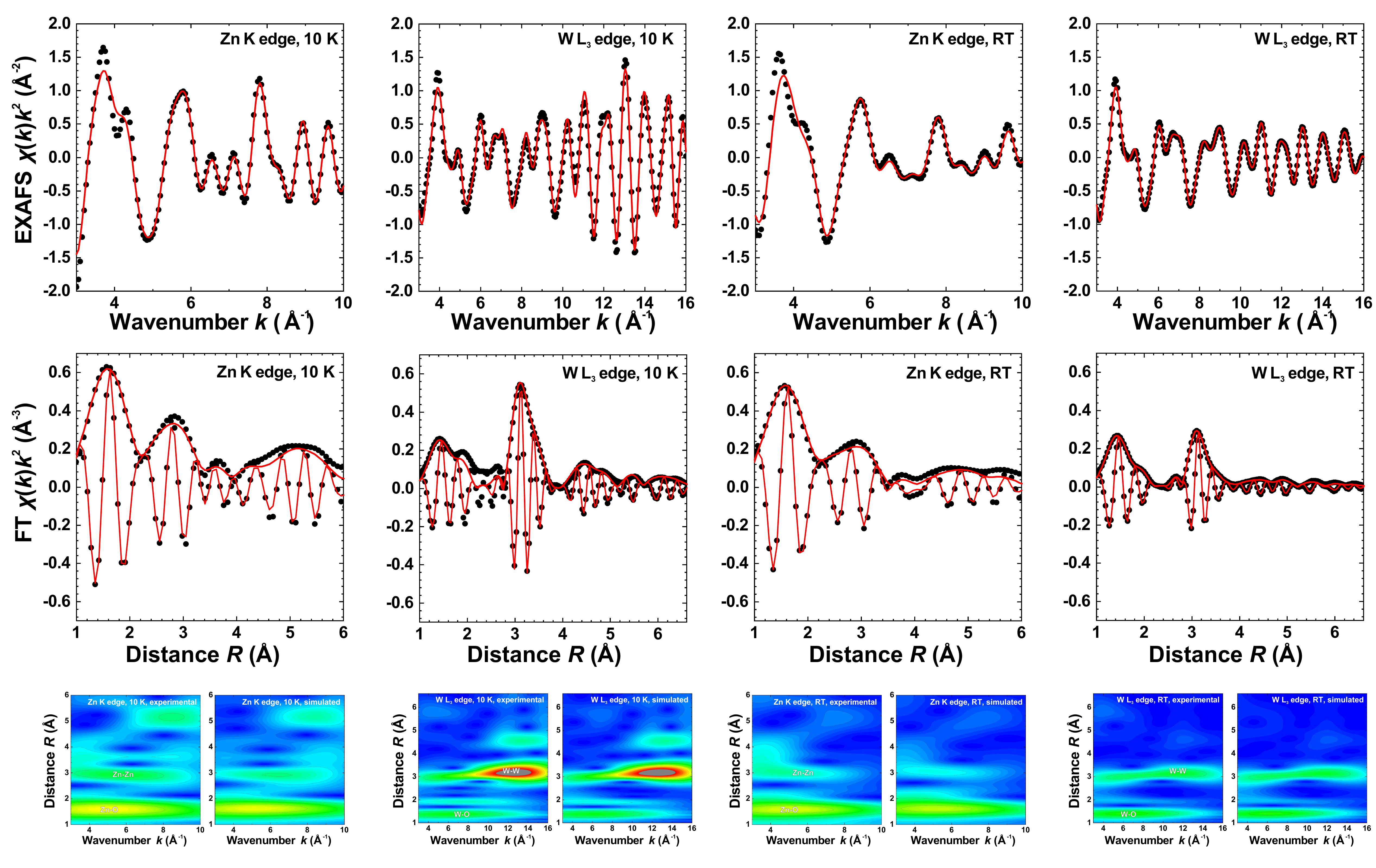}	
	\caption{Experimental (black dots) and RMC/EA calculated (solid lines) EXAFS spectra $\chi(k)k^2$ of pure \ce{ZnWO4} at the Zn K-edges and W L$_3$-edge (upper row), their Fourier transforms (FTs) (middle row) and wavelet transforms (lower rows) at 10~K and 300~K. Both modulus and imaginary parts are shown in FTs.}
	\label{figa2}
\end{figure*}

The results of the RMC/EA simulations for pure \ce{NiWO4} and \ce{ZnWO4} at 10~K and 300~K are shown in Figs.\ \ref{figa1} and \ref{figa2}, respectively.

\end{document}